# Magnetic-field-induced electronic instability of Weyl-like fermions in compressed black phosphorus

Lixuan Zheng[1†], Kaifa Luo[2,3,10†], Zeliang Sun[1†], Dan Zhao[1], Jian Li[1], Dianwu Song[1], Shunjiao Li[1], Baolei Kang[1], Linpeng Nie[1], Min Shan[1], Zhimian Wu[1], Yanbing Zhou[1], Xi Dai[3], Hongming Weng[4,5,9], Rui Yu[2*], Tao Wu[1,6,8*], and Xianhui Chen[1,6,7,8*]

[1] *CAS Key Laboratory of Strongly-coupled Quantum Matter Physics, Department of Physics, University of Science and Technology of China, Hefei 230026, China;*
[2] *School of Physics and Technology, Wuhan University, Wuhan 430072, China;*
[3] *Department of Physics, Hong Kong University of Science and Technology, Hong Kong, China;*
[4] *Beijing National Laboratory for Condensed Matter Physics, and Institute of Physics, Chinese Academy of Sciences, Beijing 100190, China;*
[5] *University of Chinese Academy of Sciences, Beijing 100049, China;*
[6] *CAS Center for Excellence in Superconducting Electronics (CENSE), Shanghai 200050, China;*
[7] *CAS Center for Excellence in Quantum Information and Quantum Physics, Hefei 230026, China;*
[8] *Collaborative Innovation Center of Advanced Microstructures, Nanjing University, Nanjing 210093, China;*
[9] *Songshan Lake Materials Laboratory, Dongguan 523808, China;*
[10] *Department of Physics, The University of Texas at Austin, Austin 78712, USA*

Revealing the role of Coulomb interaction in topological semimetals with Dirac/Weyl-like band dispersion shapes a new frontier in condensed matter physics. Topological node-line semimetals (TNLSMs), anticipated as a fertile ground for exploring electronic correlation effects due to the anisotropy associated with their node-line structure, have recently attracted considerable attention. In this study, we report an experimental observation for correlation effects in TNLSMs realized by black phosphorus (BP) under hydrostatic pressure. By performing a combination of nuclear magnetic resonance measurements and band calculations on compressed BP, a magnetic-field-induced electronic instability of Weyl-like fermions is identified under an external magnetic field parallel to the so-called nodal ring in the reciprocal space. Anomalous spin fluctuations serving as the fingerprint of electronic instability are observed at low temperatures, and they are observed to maximize at approximately 1.0 GPa. This study presents compressed BP as a realistic material platform for exploring the rich physics in strongly coupled Weyl-like fermions.

*Corresponding authors (Rui Yu, email: yurui@whu.edu.cn; Tao Wu, email: wutao@ustc.edu.cn; Xianhui Chen, email: chenxh@ustc.edu.cn)
†These authors contributed equally to this work.

# 1 Introduction

In conventional correlated metals, the Coulomb interaction between nonrelativistic quasiparticle excitations always suffers the so-called Thomas-Fermi screening effect and becomes short-ranged, which guarantees a Fermi liquid state even in the case of a moderate Coulomb interaction. However, in three-dimensional (3D) Dirac and Weyl semimetals, in which quasiparticle excitations obey linear Dirac/Weyl-like dispersion, the Coulomb interaction remains a considerable long-range component that is unscreened at the band-crossing points owing to the vanishing density of states (DOS) at the Fermi energy $E_F$. The coupling of Dirac/Weyl-like excitations with an unscreened, long-range Coulomb interaction could trigger many intriguing phenomena beyond correlation effects in conventional correlated metals, such as logarithmic velocity renormalization and Dirac cone reshaping [1-3]. For a sufficiently strong coupling with the long-range Coulomb interaction, many exotic phenomena associated with electronic instabilities of Dirac/Weyl-like fermions, such as charge density waves (CDWs) and axion or exciton insulators, could be realized above a critical coupling strength by the opening of a mass gap [4-13]. Unfortunately, in contrast to conventional correlated metals, the material realization of strongly coupled Dirac/Weyl-like fermions is fairly challenging in 3D Dirac/Weyl semimetals. Strongly coupled Weyl-like fermions were claimed in a two-dimensional (2D) organic conductor in a nuclear magnetic resonance (NMR) study [14]. Recently, the discovery of topological node-line semimetals (TNLSMs) has introduced us to a novel material family to realize strongly coupled Weyl-like fermions. One possible advantage of TNLSMs toward achieving strong correlations is rooted in the anisotropic dispersion associated with their node-line structure, which might effectively enhance the coupling strength along the node-line direction [15]. Interestingly, many recent experiments have revealed strong correlations in TNLSMs, such as ZrSiS and ZrSiSe [15-17]. Identifying whether a strong correlation drives the electronic instability of Weyl-like fermions in TNLSMs is a tremendously interesting and highly discussed topic.

In this study, we report unambiguous evidence for strongly coupled Weyl-like fermions in TNLSMs, which are realized in compressed black phosphorus (BP). Noteworthily, BP under ambient pressure is a semiconductor with a direct band gap, which has drawn extensive interest as a highly promising candidate for nanoelectronics and optoelectronics applications when exfoliated to be an atomic-thick, 2D semiconductor [18,19]. In a recent quantum oscillation experiment on bulk BP, a topological phase transition from semiconductor to topological semimetal was observed upon hydrostatic pressure application with a critical pressure $P_c$ of approximately 1.2 GPa [20,21]. Subsequently, at more than $P_c$, a first-principles calculation identified the detailed band structure of compressed BP as that of a TNLSM with a closed topological node line [22]. A high-field quantum transport measurement further indicated that possible electronic instability of TNLSMs in compressed BP might be achieved under a high magnetic field beyond the quantum limit [23]. In this study, we measure the nuclear spin-lattice relaxation rate $1/T_1$ of $^{31}$P nuclei in compressed BP in a bid to provide definite evidence for field-dependent anomalous spin fluctuations around the pressure-induced topological phase transition, which only appears under an external field parallel to the so-called nodal ring in the reciprocal space. Our band calculations further ascertain that such field-dependent anomalous spin fluctuations are closely relevant to quasi-one-dimensional (quasi-1D) Weyl-like band dispersion induced via Landau quantization.

# 2 Methods

## 2.1 Synthesis and characterization

High-quality BP single crystals were synthesized using the high-pressure synthesis technique. The growth parameter was the same as that in a previous study [20]. We selected three pieces of high-quality single crystals for NMR measurements and electrical transport measurements at hydrostatic pressures. Sample A was used for pressure-dependent NMR measurements under an applied magnetic field parallel to the $c$-axis. Sample B was cut into two pieces for pressure-dependent NMR and electrical transport measurements under an applied magnetic field parallel to the $a$-axis. Sample C was used for NMR measurements under an applied magnetic field parallel to the $c$-axis at ambient pressure. The definition of the $a$-axis, $b$-axis, and $c$-axis is shown in Figure 1(a). A representative X-ray diffraction pattern for sample A is shown in Figure S1 (Supporting Information). A set of diffraction peaks with Miller indices (0 0 $l$) indicates the high purity of a sample. Samples A and B were also characterized by employing in-plane resistance $\rho_{xx}$ and Hall resistance $\rho_{xy}$ at ambient pressure. As shown in Figure S2(a), $\rho_{xx}$ exhibits a classical behavior for doped semiconductors in both samples, consistent with previous studies [20,21]. The carrier concentration at 300 K is $1.20\times10^{17}$ and $3.08\times10^{16}$ cm$^{-3}$ for samples A and B, respectively (see Figure S2(b)). These different carrier concentrations are related to the doping level due to impurities in the samples, and they do not affect the main conclusion of this study. The temperature-dependent $1/T_1$ for all three samples also exhibits similar temperature-dependent behavior with a remarkable peak at low temperatures, as shown in Figures S2(c) and S3(a). All the abovementioned measurements demonstrate that the three samples have almost the same physical properties. Hence, our NMR measurements of all samples reveal a universal

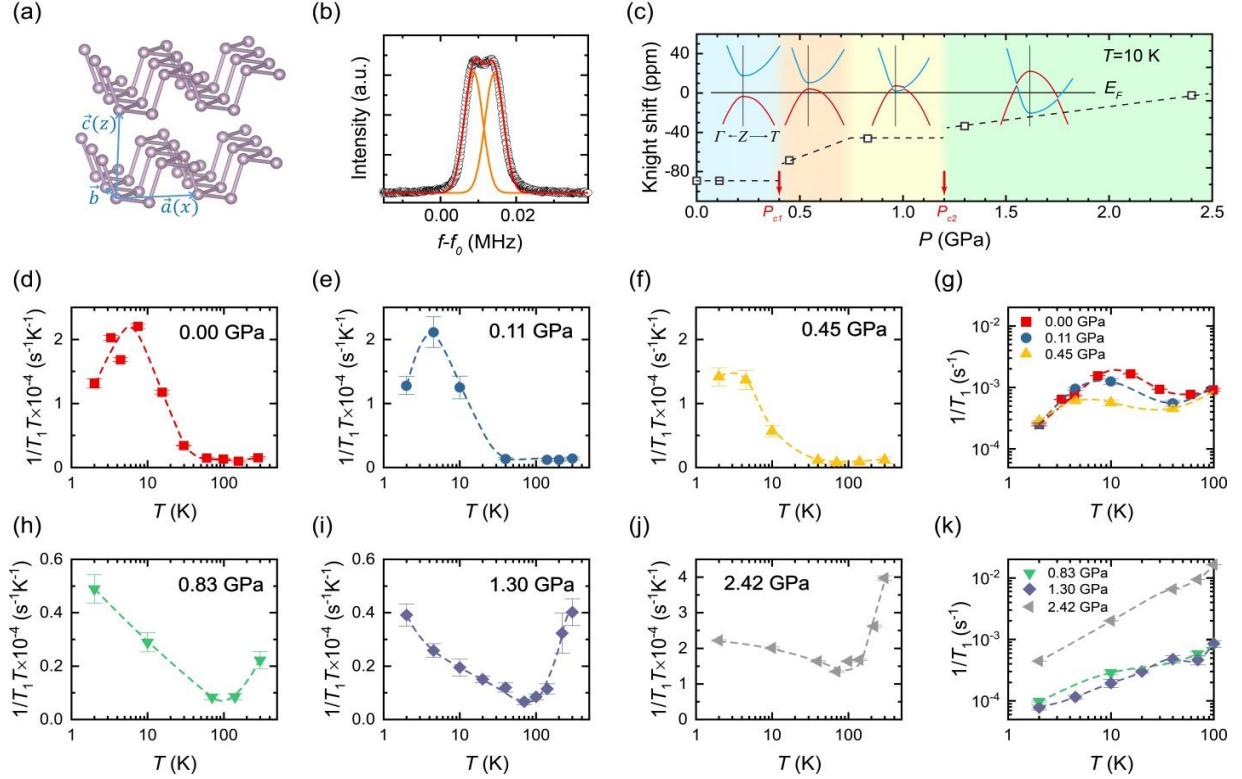

**Figure 1** (Color online) Pressure-driven topological phase transition revealed by the temperature-dependent $1/(T_1T)$. (a) Crystal structure of BP. (b) Typical $^{31}$P NMR spectrum measured at $T= 2$ K under $H = 9.8$ T along the $c$-axis. The splitting results from the dipolar field between adjacent phosphorus nuclei. (c) Two Lifshitz transitions at $P_{c1}$ (~ 0.4 GPa) and $P_{c2}$ (~ 1.2 GPa), indicated by sudden jumps in Knight shift at 10 K. Additionally, another topological phase transition occurs at approximately 0.73 GPa [21]. The insert is a sketch of the pressure-dependent evolution of the band structure around the $Z$ point in the momentum space. The colorful background represents different electronic phases with various band structures or Fermi surface topology. Red arrows are used to mark the positions of the two Lifshitz transitions. (d)-(f) Temperature-dependent $1/(T_1T)$ at pressures below 0.73 GPa. All plots are on the same scale. In this pressure range, $1/(T_1T)$ exhibits an almost temperature-independent behavior above 70 K and also a peak behavior at low temperatures. (g) Temperature-dependent $1/T_1$ at pressures below 0.73 GPa. The remarkable peak in $1/T_1$ is a hallmark of impurity-dominated nuclear spin relaxation. (h)-(j) Temperature-dependent $1/(T_1T)$ at pressures above 0.73 GPa. $1/(T_1T)$ at all pressures exhibits a similar decreasing behavior at more than 100 K and then a persistent increasing behavior below 100 K. (k) Temperature-dependent $1/T_1$ at pressures above 0.73 GPa. No peak behavior as those in (g) is observed at low temperatures. $P_{c1}$ (~ 0.4 GPa) is a critical pressure between the two fairly different low-temperature behaviors of $1/T_1$. All colorful lines in (d)-(k) are guides to the eye. $1/(T_1T)$ data above 100 K suffer the pressure loss effect mentioned in the main text. The error bars of $1/T_1$ and $1/(T_1T)$ were calculated using the error transfer formula based on the fitting error bar of $T_1$ and the measurement error bar of temperature. All data were measured under $H = 9.8$ T along the $c$-axis.

phenomenon rather than a sample-dependent one.

## 2.2 Applying hydrostatic pressure and pressure calibration

Hydrostatic pressures up to 2.42 GPa were obtained using a self-clamped, BeCu piston-cylinder cell. Daphne 7373 oil was used as a pressure-transmitting medium. We used two different calibration methods, namely, NMR and electrical transport measurements. In the NMR measurement, the nuclear quadrupole resonance (NQR) frequency of $Cu_2O$ powders was used for pressure calibration [24]. An NQR coil containing $Cu_2O$ powders was placed into the above-mentioned pressure cell together with the NMR coil. Noteworthily, the actual hydrostatic pressure always decreases with decreasing temperature owing to the freezing of Daphne 7373 [25-27]. Temperature-dependent pressure values in the pressure cell were detected at different initial pressures, as shown in Figure S4(a). The temperature-dependent pressure value was almost unchanged at less than 100 K. Therefore, the temperature-dependent NMR behavior at less than 100 K was not affected by the pressure loss effect. Additionally, a linear behavior between the calibrated pressures was confirmed in our pressure cell at temperatures both high (100 K/ 300 K) and low (2 K), as shown in Figure S4(b). The pressure at 2 K was defined as the standard pressure. The pressure calibration method is described as follows. First, the pressure was calibrated at 300 K by measuring the NQR frequency of $Cu_2O$. Second, the pressure at 2 K was obtained using the following relationship derived from Figure S4(b): $P(300 \text{ K}) = 0.934 \times P(2 \text{ K}) + 0.277$.

Here, the unit of $P$ is GPa. The uncertainty of this method is approximately 0.02 GPa.

In the electrical transport measurement, the superconducting transition temperature $T_c$ of pure tin (Sn) was used for pressure calibration [20] using the following calibration function:

$P = 2.045 \times [T_c(0) - T_c(P)]$.

The value of $T_c(0)$ is 3.72 K. The unit of $P$ is GPa. The uncertainty of this method is approximately 0.03 GPa.

### 2.3 Nuclear magnetic resonance, electrical transport, and magnetic susceptibility measurements

A commercial NMR spectrometer (Thamway Co. Ltd.) was used for the NMR measurements. A superconducting magnet (Oxford Instruments) offered a uniform magnetic field up to 16 T. Aluminum thin flake was used for calibrating the external magnetic field with uncertainty at a few ppm. All $^{31}$P NMR spectra were obtained using a standard free induction decay method. Nuclear spin-lattice relaxation time $T_1$ was measured using a saturation pulse method. Specifically, $T_1$ was obtained by fitting the recovery curve with a standard relaxation function: $1 - \frac{M(t)}{M(\infty)} = I_0 e^{-(t/T_1)}$. The error bar of $T_1$ was determined using the least-squares method. The temperature-dependent $1/T_1$ at different pressures was measured under the $H//a$-axis or $H//c$-axis with a magnitude of 9.8 T. The corresponding NMR resonant frequency was set to 168.9 MHz.

A standard dc four-probe method was performed with the physical property measurement system (PPMS-9T, quantum design). Hydrostatic pressure up to 2 GPa was applied using the self-clamped, BeCu piston-cylinder cell. The magnetic field was applied along the $a$-axis. The electrical resistance was determined from the $V$-$I$ curve with the scanning current within several mA. The magnetic susceptibility was measured using a vibrating sample magnetometer (VSM-7T, quantum design). The magnetic field applied to the sample was perpendicular to the $c$-axis and up to 7 T.

### 2.4 Electronic band structure calculation method

All calculations were performed using density-functional theory, pseudopotentials, and plane wave basis sets as implemented in the Quantum ESPRESSO suite [28,29]. We employed the generalized gradient approximation of exchange-correlation functional in Perdew-Burke-Ernzerhof [30] type and optimized norm-conserving pseudopotentials [31] from the PseudoDojo library [32] for total energy. The plane wave kinetic energy cutoff was set to 44 Rydberg (600 eV) to achieve 1 meV accuracy as recommended in the library, and Brillouin zone integration was performed on a shifted mesh of 14×6×10 $k$-points. Near the topological phase transition driven by hydrostatic pressure, the energy gap at $Z$ point was almost linearly dependent on unit-cell volume compression [22]. We selected the band structure at 1.4 GPa as a representative case of topological nodal ring. The unit-cell volume is 75.81 Å$^3$. The lattice parameters are $a$ = 3.31 Å ($x$-direction), $b$ = 10.47 Å ($z$-direction), and $c$ = 4.37 Å ($y$-direction). Additionally, four phosphorus atomic positions are ($x, y, z$), ($-x, -y + 1, -z + 1$), ($x + 1/2, y + 1/2, -z + 1/2$), and ($-x - 1/2, -y + 1/2, z + 1/2$) where $x = -0.103$, $y = 0.103$, $z = 0.079$ in unit of crystal vector.

## 3 Experimental results

With an aim to conduct high-pressure NMR measurements under the external magnetic field ($H$) along different crystalline axes, we selected two pieces of high-quality BP single crystals (samples A and B) synthesized using the high-pressure synthesis technique with almost the same physical properties at ambient pressure. The crystalline axes are defined in Figure 1(a). Furthermore, samples A and B were used for high-pressure NMR measurements under the $H//c$-axis and $H//a$-axis, respectively. A representative NMR spectrum of $^{31}$P nuclei is shown in Figure 1(b), which exhibits only one kind of $^{31}$P site with a tiny splitting resulting from the mutual magnetic dipolar interactions among adjacent $^{31}$P nuclear spins [33,34]. Based on a previous quantum oscillation experimental study [20-21], we can say that two Lifshitz transitions should occur upon increasing the pressure. Upon increasing the pressure, a single hole-like pocket was observed at more than $P_{c1}$, which indicates the impurity-pinned Fermi level across the valence band at $P_{c1}$. With further increasing pressure, the residual semiconducting gap above $E_F$ closed at approximately 0.73 GPa [21], following which a node-line semimetal phase with both hole- and electron-like pockets at $E_F$ appeared through another Lifshitz transition at approximately $P_{c2}$ [20]. The above-mentioned pressure-induced Lifshitz transitions in BP are shown in the inset of Figure 1(c). Usually, the Knight shift, which is relative to local magnetic susceptibility, is sensitive to Lifshitz transitions and would exhibit a jump-like behavior. Two jump-like behaviors at $P_{c1}$ (~ 0.4 GPa) and $P_{c2}$ (~ 1.2 GPa) were confirmed through our measurement, as shown in Figure 1(c). Next, we systematically studied the pressure-dependent evolution of $1/(T_1T)$ at different pressure ranges divided by these two Lifshitz transitions. The values of the temperature-dependent $1/(T_1T)$ under the $H//c$-axis at pressures up to 2.42 GPa are shown in Figure 1(d)-(f) and (h)-(j). The most important finding is the unusual behavior of $1/(T_1T)$ at low temperatures.

The temperature-dependent $1/(T_1T)$ in different phases also exhibits distinct temperature-dependent behaviors as temperature decreases from room temperature. As explained in the Methods section, the pressure in our pressure cell always decreases with cooling and becomes stable only at less than 100 K. Therefore, only the experimental results obtained at less than 100 K were quantitatively analyzed in the following. At less than 0.73 GPa, the temperature-dependent $1/T_1$ exhibits a remarkable peak behavior at low temperatures, as shown in Figure 1(g). In sharp contrast, at more than 0.73 GPa, a monotonic decrease behavior is observed at low temperatures (see Figure 1(k)). Such two distinct behaviors indicate that a certain transition occurs at approximately 0.73 GPa. In the semiconducting phase (corresponding to Figure 1(g) at less than $P_{c1}$), there are two kinds of relaxation channels in a nuclear spin relaxation process, namely, thermally activated carriers and randomly fluctuating magnetic fields. The former originates from the energy level of impurities pinned inside the intrinsic band gap. The latter is probably induced by localized spins in impurities (see more discussion in Supporting Information S2). Because the contribution from thermally activated carriers on $1/T_1$ becomes negligible at low temperatures, the low-temperature, peak-like behavior should be ascribed to the magnetic fluctuations induced by localized spins [35-38]. Because the local spins are randomly distributed in the sample, we can use the Bloembergen-Purcell-Pound (BPP) model, which is appropriate for incoherent local spins, to describe the nuclear spin relaxation [39]. In the BPP model, the nuclear spin relaxation rate is expressed as follows:

$$\frac{1}{T_1} = \gamma_n^2 \langle h_\perp \rangle^2 \frac{\tau_c}{1+\omega^2 \tau_c^2}, \quad (1)$$

where $\gamma_n$ represents the gyromagnetic ratio of measured nuclei, $h_\perp$ the transverse magnetic hyperfine field due to local moments, $\tau_c$ the correlation time of local moments, and $\omega$ the Larmor frequency. If only considering a temperature-dependent $\tau_c$, which monotonically increases with cooling, then eq. (1) would give a peak-like behavior of $1/T_1$, and also the maximum $1/T_1$ must satisfy the condition $\tau_c = 1/\omega$. Moreover, the BPP model also gives an $H^2$-dependent $T_1$ below the peak temperature and an $H$-independent $T_1$ above the peak temperature. This is also confirmed in our experiment, which further supports a BPP process due to local spins (see Figures 2(d) and S3(b)). The concentration of local spins was estimated to be approximately $3\times10^5$ mol$^{-1}$ through magnetic susceptibility measurements, which also satisfies the prerequisite for the BPP model [39] (for details, see Supporting Information S2). Therefore, we conclude that the low-temperature nuclear spin-lattice relaxation in the semiconducting phase below $P_{c1}$ is dominated by local spins in impurities. Noteworthily, the local spins in impurities originate from the localized unpaired electrons at the in-gap impurity levels, which would be delocalized with increasing temperature or closing the band gap (see details in Supporting Information S2).

When the applied hydrostatic pressure is beyond 0.73 GPa (i.e., $> P_{c2}$), the semiconducting gap above $E_F$ is completely closed, and the quasiparticle excitations at the Fermi level rather than local spins should become the dominant channel for the nuclear spin-lattice relaxation process. Three pieces of evidence support this assertion. First, the peak-like behavior due to local spins completely disappears at low temperatures in $1/T_1$, as shown in Figure 1(g). Second, the field dependence of $T_1$ deviates from $H^2$ above 0.73 GPa, as shown in Figure 2(e) (for details, see Supporting Information S2). Third, the pressure-dependent, fluctuating magnetic field extracted from the BPP model linearly decreases to zero at approximately 0.7 GPa (for details, see Supporting Information S2). All these facts strongly suggest a dominant relaxation channel from the quasiparticle excitations at $E_F$. As previously discussed, the band structure above $P_{c2}$ is identified as that of a TNLSM with a closed topological node line across $E_F$. An approximate Weyl-like dispersion around $E_F$ would be naturally expected to govern the nuclear spin relaxation. Earlier theoretical and experimental studies in Dirac or Weyl semimetals [40-43] revealed a distinct nuclear spin relaxation process beyond a Fermi liquid with nonrelativistic quasiparticles, which originates from an unusual orbital hyperfine coupling tensor and special energy dependence of DOS. Whether a similar nuclear spin relaxation behavior is applicable to TNLSMs remains to be elusive. Although the accuracy of the temperature-dependent $1/(T_1 T)$ at more than 100 K is affected by the abovementioned pressure loss effect, the possible behavior of $1/T_1$ in TNLSMs has been discussed in Supporting Information S3. The focus of this NMR study is not $1/(T_1 T)$ behavior above 100 K but a divergent behavior at low temperatures under the $H//c$-axis, which is not affected by the pressure loss effect above 100 K. The temperature-dependent $1/(T_1 T)$ exhibits a remarkable upturn behavior below 100 K, as shown in Figure 2(a). The fitting result indicates that the divergent behavior at low temperatures follows a power-law behavior with $n = -1/2$. In stark contrast, the temperature-dependent $1/(T_1 T)$ under the $H//a$-axis exhibits a moderate enhancement below 100 K and finally saturates at the lower temperatures. Recently, a slight upturn behavior due to a temperature-dependent chemical potential $\mu(T)$ was observed in topological semimetals [44]. The temperature dependence of $\mu(T)$ is attributed to the small Fermi energy comparable to $k_B T$ in topological semimetals. In BP, the Fermi energy is also considerably small, i.e., ~ 11.8 meV, near the critical pressure $P_{c2}$ [23]. Therefore, a similar upturn behavior is also expected at low temperatures, and we observed exactly this under the $H//a$-axis (for more details, see Figure 2(a) and Supporting Information S3). However, noteworthily, the $\mu(T)$-induced upturn behavior would finally saturate at lower temperatures [44], which can well explain the $1/(T_1 T)$ behavior under the $H//a$-axis but not $1/(T_1 T)$ behavior under the $H//c$-axis (see Figure 2(a)). Therefore, the noninteracting Dirac/Weyl fermions cannot account for the low-temperature persistent divergent behavior in $1/(T_1 T)$ under the $H//c$-axis.

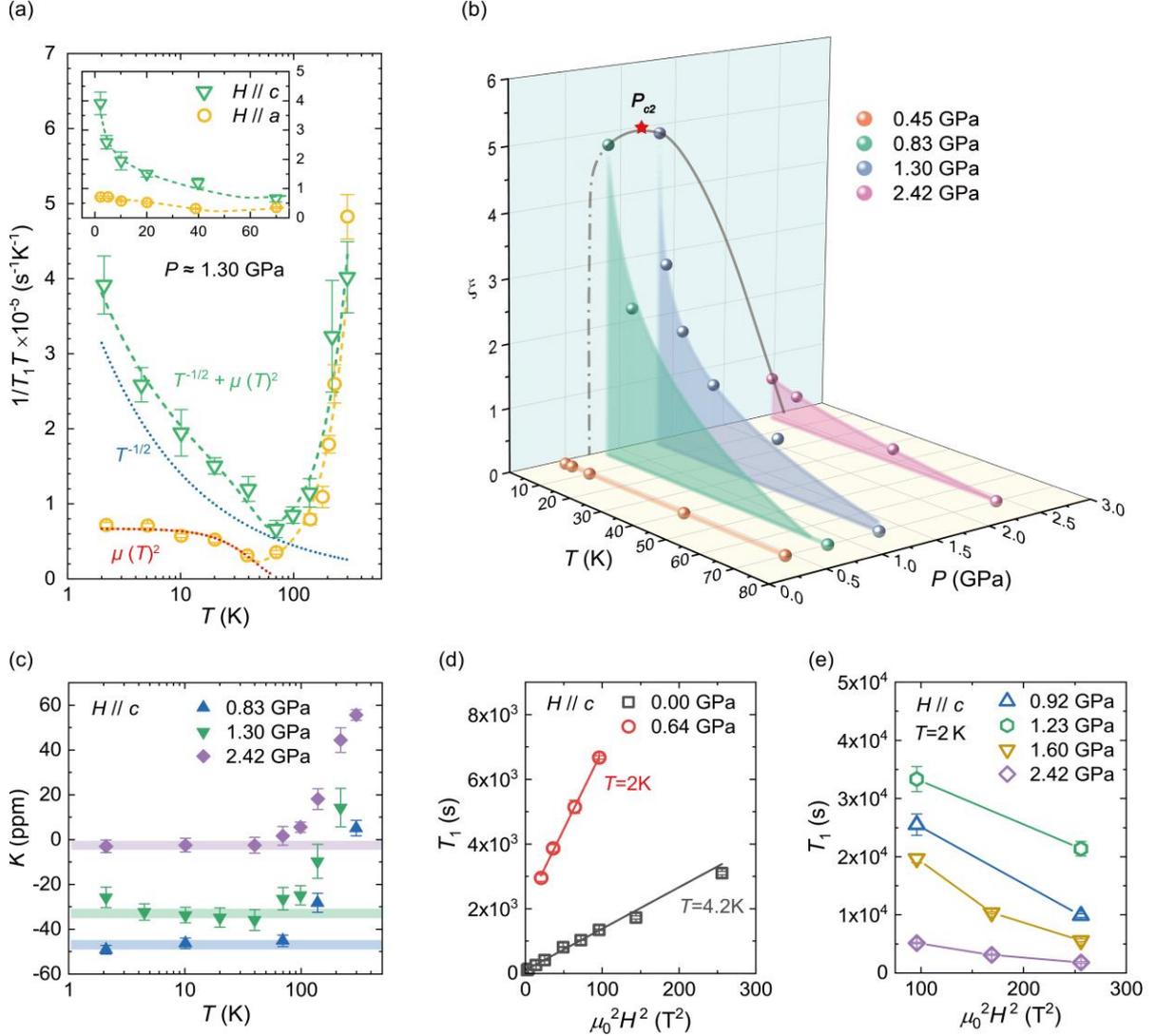

**Figure 2** (Color online) Evidence of anomalous spin fluctuations in the TNLSM state. (a) Anisotropic $1/(T_1T)$ at approximately 1.30 GPa under $H = 9.8$ T. According to the band structure calculation, $1/(T_1T)$ plotted by green (yellow) open points is obtained for the Weyl-like (completely gapped) band structure under the $H//c$ ($a$)-axis. The dashed lines represent fitting $1/(T_1T)$ curves (for details, see Supporting Information S3). The blue dotted line represents the power-law component in the fitting curve. The red dotted line represents the component due to $\mu(T)$ in the total fitting curve. $1/(T_1T)$ under the $H//c$-axis can befitted by the sum of the two abovementioned components. In contrast, $1/(T_1T)$ under the $H//a$-axis only needs a single component due to $\mu(T)$. The inset is $1/(T_1T)$ plotted in a linear scale below 70 K. (b) Pressure dependence of spin fluctuation strength under $H = 9.8$ T along the $c$-axis. $\xi$ represents a dimensionless parameter to scale spin fluctuation strength (for details, see Supporting Information S4). Because the impurity-induced magnetic fluctuations dominate the nuclear spin relaxation at less than 0.64 GPa, $\xi$ ought to be 0 at 0.45 GPa according to the definition. Colorful lines represent $T^{-1/2}$ power-law fitting. The gray dash-dot line is a guide to eyes for the spin fluctuation strength at 2 K. (c) Temperature-dependent Knight shift above 0.73 GPa under the $H//c$-axis. The Knight shift data above 100 K suffer the pressure loss effect mentioned in the main text. The error bar of Knight shift is determined from the Gaussian fitting of the NMR spectra. (d) Field-dependent $T_1$ in the semiconductor phase under the $H//c$-axis. Data at 0 and 0.64 GPa were obtained at 4.2 and 2 K, respectively. (e) Field-dependent $T_1$ in the TNLSM state under the $H//c$-axis. The positive linear relationship between $T_1$ and $H^2$ absolutely breaks down.

What is the physical origin for the persistent enhancement of $1/(T_1T)$ at low temperatures? How about the local spins? One possibility is to consider the Kondo coupling between itinerant electrons and local spins, in which $1/(T_1T)$ could be enhanced at a temperature exceeding the Kondo temperature $T_K$ owing to incoherent Kondo scattering. However, in this case, $1/(T_1T)$ should also be suppressed by increasing the applied magnetic field [45], which is inconsistent with the field-dependent results shown in Figure 2(e). Moreover, our analysis on $1/T_1$ indicates that the magnetic fluctuations from local spins disappear at approximately 0.7 GPa, coinciding with the closure of the band gap (see Supporting Information S2). All these facts strongly support that the role of local spins on the persistent enhancement of $1/(T_1T)$ could be eliminated. Recently, a disorder-driven quantum critical point was also proposed in Weyl semimetals, which would enhance the upturn behavior in $1/(T_1T)$ owing to $\mu(T)$ [46]. However, even in this situation, the temperature dependence of $1/(T_1T)$ would become constant at low temperatures rather than exhibiting divergent behavior. Finally, we emphasize that the divergent behavior in $1/(T_1T)$ strongly depends on the orientation of the magnetic field, which only emerges for the $H//c$-axis but not for the $H//a$-axis. To the best of our knowledge, such a highly anisotropic behavior in $1/(T_1T)$ should not be ascribed to any known impurity-related effect. Therefore, we conclude that the divergent behavior observed in $1/(T_1T)$ should be an intrinsic effect in the TNLSM state.

A plausible explanation of the persistent enhancement of $1/(T_1T)$ at low temperatures is the critical spin fluctuations due to certain electronic instabilities of interacting Weyl-like fermions. Recently, a similar low-temperature upturn behavior in $1/(T_1T)$ was observed for an organic conductor with interacting 2D Weyl fermions, which was ascribed to internode excitonic fluctuations [14]. In principle, the electronic excitations around Weyl nodes can be categorized into two processes characterized by contrasting momentum transfers ($\hbar\mathbf{Q}$): $C_{\mathbf{Q}\sim 0}$ and $C_{\mathbf{Q}\sim 2k}$. Usually, $1/(T_1T)$ probes the sum of $C_{\mathbf{Q}\sim 0}$ and $C_{\mathbf{Q}\sim 2k}$, whereas Knight shift ($K$) examines only $C_{\mathbf{Q}\sim 0}$ (in particular, $\mathbf{Q} = 0$). In contrast to $1/(T_1T)$, Knight shift becomes constant at low temperatures rather than exhibiting any divergent behavior (see Figure 2(c)), which strongly suggests that the critical spin fluctuations are dominated by the $C_{\mathbf{Q}\sim 2k}$ process between Weyl nodes. In order to map out the pressure dependence of the $C_{\mathbf{Q}\sim 2k}$ process between Weyl nodes, we define a dimensionless parameter $\xi$ to scale temperature-dependent spin fluctuations at different pressures:

$$\xi = \frac{[1/(T_1T)] - [1/(T_1T)]^{\min}}{[1/(T_1T)]^{\min}},$$

where $[1/(T_1T)]^{\min}$ denotes the minimum value for each $1/(T_1T)$ curve at different pressures. By only considering the contribution from $C_{\mathbf{Q}\sim 0}$ and $C_{\mathbf{Q}\sim 2k}$, the expression for $\xi$ can be simplified to $\xi \sim C_{\mathbf{Q}\sim 2k}/C_{\mathbf{Q}\sim 0}$ (for details, see Supporting Information S4). Accordingly, the disappearance of the divergent behavior in $1/(T_1T)$ under the $H//a$-axis implies that the $C_{\mathbf{Q}\sim 2k}$ process should be strongly suppressed compared with that under the $H//c$-axis.

## 4 Discussion on band structure and Fermi surface instability

We calculated the band structure of compressed BP under different magnetic fields to further understand the above-mentioned anisotropic behavior in $1/(T_1T)$. We observed that the formation of a quasi-1D Weyl-like band structure under magnetic fields was the key to understanding the confounding behavior of $1/(T_1T)$. Compressed BP exhibited a band structure near the Fermi level characterized by a conduction band and valence band. These bands intersect to form a continuous nodal ring within the momentum space in the TNLSM state [22], as shown in Figure 3(a). There are two possible topological invariants labeling the TNLSM state in compressed BP: 1D Berry phase and the first Chern number [47]. In compressed BP, the former can be nontrivial, while the latter is always trivial. From first-principles calculations, we constructed a symmetry-constrained minimal effective model to capture the low-energy physics of our interest (derivation is detailed in Supporting Information S5). Based on this effective Hamiltonian, we analyzed the band structure involving the lowest two Landau levels. As shown in Figure 3, our results indicate that the nodal ring in the band structure under zero magnetic field would evolve into a quasi-1D Weyl-like band structure under $H$ parallel to the nodal ring (corresponding to the $H//c$-axis in real space) and a completely gapped band structure under $H$ perpendicular to the nodal ring (corresponding to the $H//a$-axis in real space), respectively (see the calculation details in Supporting Information S6). The formation of distinct band structures under different magnetic field orientations can naturally explain the anisotropic behavior in $1/(T_1T)$, in which the divergent behavior under the $H//c$-axis is ascribed to Weyl node scattering induced by electronic correlations. Theoretically, the instability induced by Weyl node scattering is closely connected with the distance of Weyl nodes from the Fermi level [12]. Under the $H//c$-axis, a possible situation is that the distance between the emergent Weyl nodes and Fermi level remains unchanged until the pressure increases up to $P_{c2}$, and then it considerably changes due to the strong band inversion at more than $P_{c2}$ (see the illustration in Figure 2(c)). This explains the pressure dependence of the divergent behavior of $1/(T_1T)$. Additionally, we also consider possible Peierls instability driven by Fermi surface nesting, which strongly relies on Fermi surface topology. The Shubnikov-de Haas (SdH) quantum oscillation results under the $H//c$-axis and $H//a$-axis can offer a constraint to determine the practical Fermi level in the calculated band structure (see Supporting Information S5) and also the relevant Fermi surface topology. The Fermi level was determined to be approximately 29 meV below the degenerate points in the quasi-1D Weyl-like band structure under the $H//c$-axis. The same Fermi level was out of the energy gap in the quasi-1D electronic structure under the $H//a$-axis (see Figure S5). Therefore, considering a 1D Peierls instability-driven CDW to account for the abovementioned electronic instability, a similar divergent behavior in $1/(T_1T)$ should also be expected under the $H//a$-axis (for details, see Supporting Information S7). This is inconsistent with our NMR observation. Based on the abovementioned analysis using the minimal effective model, we conclude that the divergent behavior of $1/(T_1T)$ only under the $H//c$-axis originates from the internode electronic instability of the quasi-1D Weyl-like band structure induced via Landau quantization.

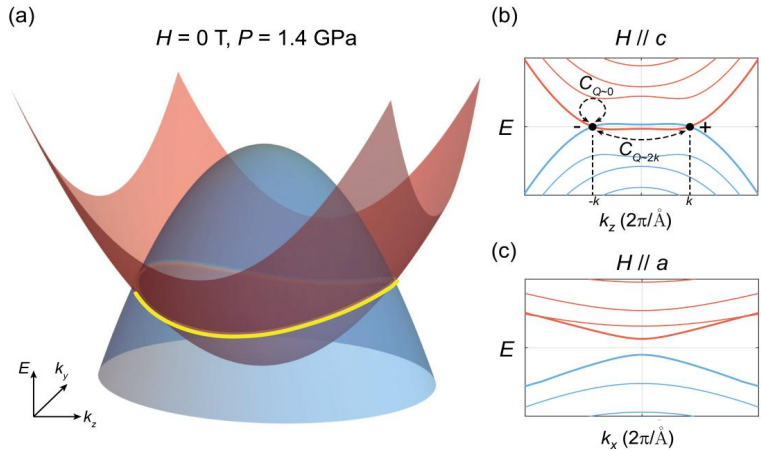

**Figure 3** (Color online) Band calculations for compressed BP at 1.4 GPa under the $H//a$-axis and $H//c$-axis. (a) Nodal ring (yellow) crossed by two bands (red and blue) from the $k \cdot p$ model. It lies in the $b$-$c$ plane of the real space. (b), (c) Numerical Landau levels of the nodal ring band structure under the $H//c$-axis and $H//a$-axis, respectively. From the effective Hamiltonian, the nodal ring transits into a Weyl-like band structure under the $H//$ring (along the $c$- or $b$-axis). The nodal ring is completely gapped under the $H \perp$ring (just along the $a$-axis). The intranode scattering process $C_{\mathbf{Q}\sim 0}$ and the internode scattering process $C_{\mathbf{Q}\sim 2k}$ are represented by black dash lines in (b). The chirality of a couple of nodes is also labeled. Details regarding the band calculation are provided in Supporting Information S5 and S6.

The electronic instability of various topological semimetals was widely discussed in theory [5-14]. A widely discussed scenario is electron-electron interaction-driven electronic instability of a Dirac/Weyl node with gap opening, which is similar to the context of chiral symmetry breaking in particle physics [48,49]. The long-range Coulomb interaction due to poor screening at a Dirac/Weyl node is frequently attributed to trigger such electronic instability [6,50]. With increasing external magnetic field, the high-energy Landau levels would be driven away from the Fermi level such that their contributions to the screening of the lowest Landau level are gradually weakened, which should enhance the abovementioned electronic instability. The field dependence of $T_1$ in the TNLSM state strongly supports the abovementioned scenario, as shown in Figure 2(e). More-over, with pressure increasing to more than $P_{c2}$, the DOS at the Fermi level is rapidly increased, which should be advantageous for screening the long-range Coulomb interaction. The pressure-dependent $\xi$ indicates that the spin fluctuations due to internode excitations are largely reduced at more than $P_{c2}$ (see Figure 2(b)), which is also consistent with the abovementioned scenario. However, considering the realistic spin-orbital coupling (SOC) in the theoretical calculation, all degenerate points of the Weyl-like band structure would open an energy gap of approximately 6 meV (for details, see Supporting Information S5). Knowing whether the long-range Coulomb interaction could still trigger similar electronic instability in this case remains elusive. Recently, a remarkable renormalization effect due to a short-range Coulomb interaction was revealed in ZrSiSe by means of optical spectroscopy [15]. Noteworthily, ZrSiSe is also a TNLSM whose degenerate points are also gapped out via SOC. This result strongly suggests a possible role of the short-range Coulomb interaction in the correlation effect in TNLSMs. This might also be applicable in compressed BP, but further theoretical investigation is required to clarify the origin of the correlation effect.

## 5 Conclusion

We provided a phase diagram of compressed BP in Figure 4. At less than $P_{c1}$, the electronic system was in a semiconducting phase. In this region, there was no Fermi surface and observable quantum oscillation in transport measurements. The nuclear spin-lattice relaxation at low temperatures was dominated by local spins in the impurities, and the contribution from the itinerant electrons was negligible ($\xi \sim 0$). At more than $P_{c1}$, i.e., when $P < 0.73$ GPa, the semiconducting gap remained finite above the Fermi level, but the impurity-pinned Fermi level crossed over the valence band. Although there was a single Fermi pocket in the quantum oscillation experiment, the nuclear spin-lattice relaxation was still dominated by local spins in the impurities at low temperatures. When $P > 0.73$ GPa, the semiconducting gap above the Fermi level was completely closed, following which the electronic system evolved into the TNLSM state. Although the node ring was above the Fermi level in the region, anomalous spin fluctuations due to internode electronic excitations rather than local spins in the impurities already became predominant in $1/(T_1T)$. Finally, when $P > P_{c2}$, the nodal ring also crossed over the Fermi level, following which two kinds of Fermi pockets were observed in the quantum oscillation experiment: one electron-like pocket and one hole-like pocket. Simultaneously, the anomalous spin fluctuations gradually weakened with further increasing pressure. The magnetic field-induced, quasi-1D Weyl-like band structure was considered a key factor for such anom-alous spin fluctuations.

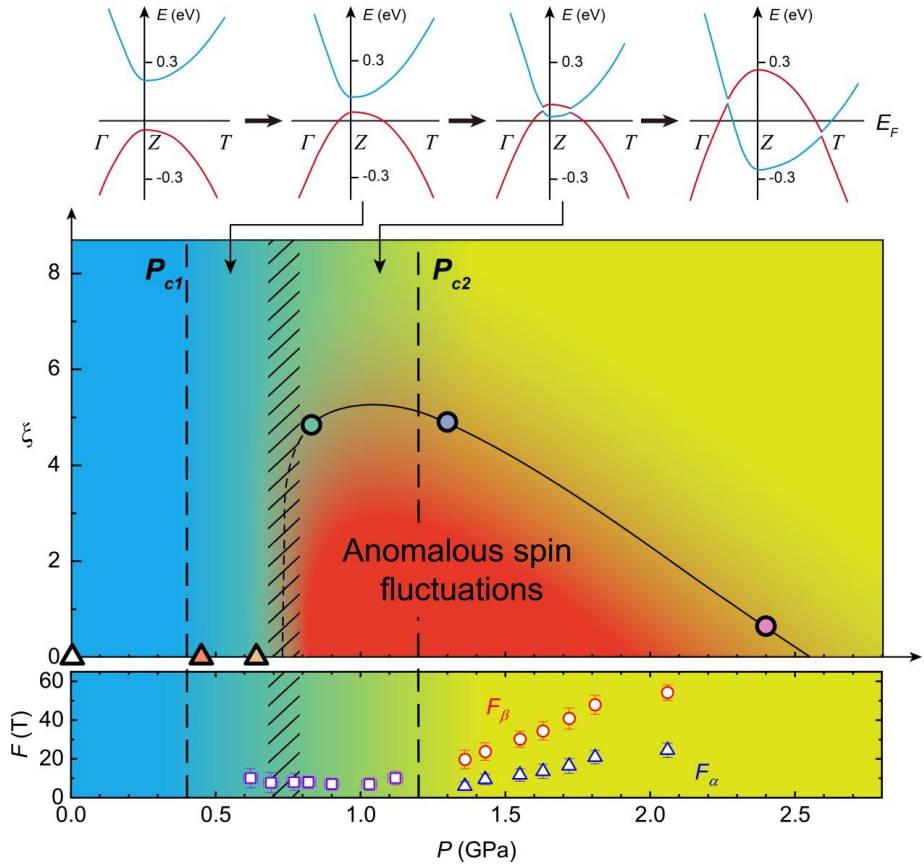

**Figure 4** (Color online) Summarized pressure-dependent phase diagram of BP. The top of the phase diagram illustrates the band structure evolution. The energy scale is an estimated value. From left to right, the band structure undergoes a Liftshitz transition at $P_{c1}$, a topological phase transition from the semiconductor state to the TNLSM state at approximately 0.73 GPa (marked by the shadow line), and then another Liftshitz transition at $P_{c2}$. The bottom plot in the phase diagram shows the pressure-dependent SdH quantum oscillation frequency obtained from Figure 3 of ref. [20]. Two Lifshitz transitions at $P_{c1}$ and $P_{c2}$ can be clearly revealed. Colorful points at the center of the phase diagram represent the dimensionless parameter $\xi$ at 2 K, which describes the strength of anomalous spin fluctuations. Interestingly, the anomalous spin fluctuations only exist above 0.73 GPa, where the TNLSM state emerges via a topological phase transition. This phase diagram shows an intimate connection between anomalous spin fluctuations and the TNLSM state. Reproduced with permission from ref. [20].

The anomalous spin fluctuations due to internode electronic excitations, which are the main findings in this study, are a hallmark of strongly correlated Weyl-like fermions. The NMR results presented TNLSMs as a promising avenue for realizing the electronic instability of interacting Weyl-like fermions. In fact, TNLSMs might be a fertile ground for exploring exotic quantum states of matter, such as topological superconducting state and axion insulating state. This study also indicated compressed BP as a realistic material platform to explore the rich physics in strongly coupled Weyl-like fermions, which would stimulate further experimental and theoretical explorations in compressed BP.

*This work was supported by the National Key R&D Program of the Ministry of Science and Technology of China (Grant Nos. 2017YFA0300201, and 2016YFA0303000), the Anhui Initiative in Quantum Information Technologies (Grant No. AHY160000), the National Natural Science Foundation of China (Grant No. 11534010), and the Key Research Program of Frontier Sciences, Chinese Academy of Sciences, China (Grant No. QYZDY-SSW-SLH021). Tao Wu thanks Yi Zhou, Zhong Wang, and Guozhu Liu for insightful discussion.*

**Conflict of interest** The authors declare that they have no conflict of interest.

**Supporting Information**

The supporting information is available online at http://phys.scichina.com and https://link.springer.com. The supporting materials are published as submitted, without typesetting or editing. The responsibility for scientific accuracy and content remains entirely with the authors.